\documentclass[aps,prl,twocolumn,letterpaper,superscriptaddress]{revtex4-2}
\usepackage[colorlinks,linkcolor=blue,anchorcolor=blue, citecolor=blue,urlcolor=blue,]{hyperref}
\usepackage{graphicx}
\usepackage{CJK}
\usepackage{verbatim}
\usepackage{braket}
\usepackage{amsmath}
\usepackage{lineno}
\usepackage{bm}
\usepackage{mathptmx}
\usepackage{subfigure}
\makeatletter

\usepackage{titlesec}
\titleformat*{\section}{\large\centering\bfseries}
\usepackage{subfiles}
\usepackage{natbib}
\usepackage{lipsum}

\begin{document}

\begin{CJK*}{UTF8}{bsmi}

\title{Showcasing the necessity of the principle of relative motion in physical statistics\\Inconsistency of the `segmented Fermi surface'}

\author{Wei Ku (\CJKfamily{bsmi}顧威)}
\altaffiliation{corresponding email: weiku@sjtu.edu.cn}
\affiliation{Tsung-Dao Lee Institute, Shanghai 200240, China}
\affiliation{Key Laboratory of Artificial Structures and Quantum Control (Ministry of Education), Shanghai 200240, China}
\affiliation{Shanghai Branch, Hefei National Laboratory, Shanghai 201315, China}
\author{Anthony Hegg}
\affiliation{Tsung-Dao Lee Institute, Shanghai 200240, China}
\affiliation{Shanghai Branch, Hefei National Laboratory, Shanghai 201315, China}

\date{\today}

\begin{abstract}
The hunt for exotic properties in flowing systems is a popular and active field of study, and has recently gained renewed attention through claims such as a ``segmented Fermi surface'' in a superconducting system that hosts steady superflow of screening current driven by an external field. Apart from this excitement and the promise of hosting Majorana zero modes, claims such as this imply exotic gap-to-gapless quantum phase transitions \textit{merely} through boost of inertial frames of observation, and challenge the very concept behind the principle of relative motion. Here, we first illustrate an obvious inescapable physical inconsistency of such claims concerning the flow velocity. Taking into account this basic principle from the beginning, we then demonstrate that a proper employment of physical statistics naturally reproduces the experimental observation without causing such a conceptual crisis. This example showcases the importance of \textit{strict} adherence to the basic principle of relative motion in physical statistics, especially when pushing the frontiers of physics and technology.
\end{abstract}
\maketitle
\end{CJK*}


The recent claim of an observed ``segmented Fermi surface'' on the surface of a fully gapped superconductor hosting a steady magnetic field-drive screening current~\cite{jfjia} has generated strong interest given its promise of hosting Majorana end states for potential application in quantum computing.
Note, however, that the claimed discovery in fact implies a insulator-metal like gapped-to-gapless quantum phase transition (from zero field to a small field) simply via effectively changing the motion of the rest frame of a  superflow.
Similar to a recent claim of quantum insulator-superconducting phase transition via Galilean transformation~\cite{ye1}, such claims have serious implications as they challenge physicists' fundamental assumption on the principle of relative motion (or its foundation concerning an observation-independent reality).

To see this serious implication, recall that a \textit{steady} superflow driven by turning on external (magnetic~\cite{london} or electric) field can be more intuitively understood from a \emph{stationary state} in its inertial rest frame~\cite{landau9} owing to its complete lack of dissipation.
This is because in this frame, the moving `environment', such as the wall of the pipe~\cite{landau9} or the crystal lattice, does not generate any dissipation to the superflow (by definition) and therefore does not directly affect the latter.
Correspondingly, the essential effect of a weak external field is simply to move the rest frame of the superflow (described by the \emph{same} quantum state) with respect to the lab frame~\cite{landau9}.
The principle of relative motion then requires that in its rest frame this state be insensitive to such a change of lab frame of the observer.
Therefore, the principle of relative motion should not allow the above phase transition or the claimed discovery.

Below we first illustrate an obvious physical inconsistency of such claims concerning the flow velocity. We then demonstrate that a proper employment of physical statistics naturally reproduces the experimental observation without causing such a conceptual crisis. This example serves as a reminder that physical statistics must \textit{strictly} adhere to the fundamental principle of relativity.

\begin{figure}[t]
\renewcommand\thefigure{\arabic{figure}}
\includegraphics[height = 3.25cm]{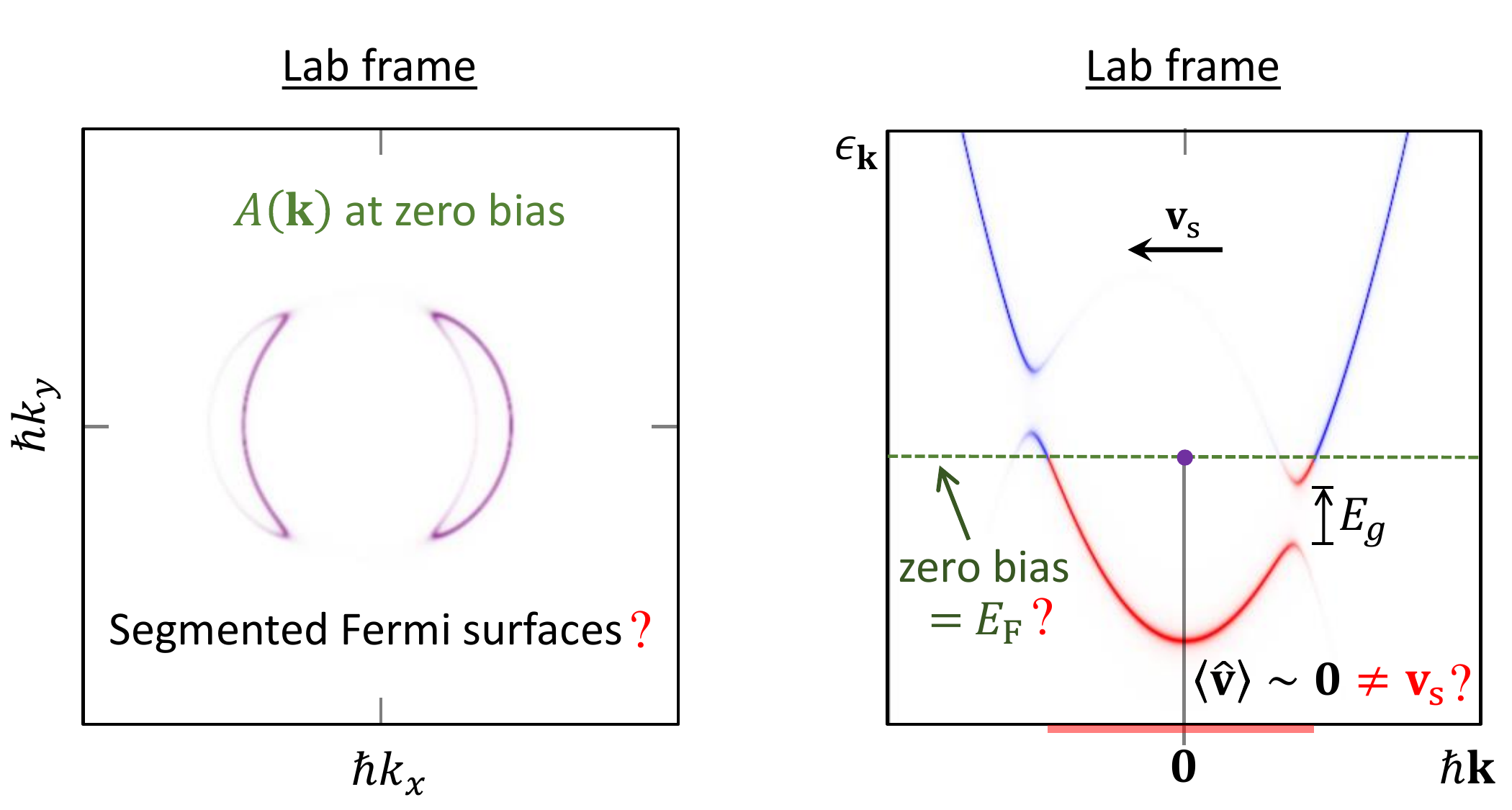}
\caption{Figure 1  Illustration of the claimed ``segmented Fermi surface'' (left panel) and the corresponding energy-momentum $(\epsilon,\hbar\mathbf{k})$-dispersion (right panel) of the one-body spectral function $A(\mathbf{k})$ for an $s$-wave superconducting surface state that supports a steadily flowing screening current driven by, for example, an external magnetic field. The occupied one-body states (in red) and the unoccupied ones (in blue) are separated by the Fermi level assumed to coincide with the zero bias voltage (in green). Due to the finite flow velocity of the screening current, the dispersion $\epsilon_\mathbf{k}$ acquires a $\mathbf{k}$-dependent shift such that for some $\mathbf{k}$ the superconductivity induced gap $E_g$ is away from the Fermi level, allowing the corresponding $\epsilon_\mathbf{k}$ to intercept with the Fermi level and form the ``segmented Fermi surfaces''.
Notice, however, that assigning the zero bias level as the Fermi energy would populate $\mathbf{k}$ and $-\mathbf{k}$ almost evenly, as denoted by the thick red line, inconsistent with the finite flow of the system.
Indeed, the average velocity $\braket{\hat{\mathbf{v}}}=(\hbar\Omega)^{-1} \int f_\mathbf{k} \mathbf{\nabla}_\mathbf{k} \epsilon_\mathbf{k} d^2 k$ of the system is basically zero, in direct contradiction to the finite velocity $\mathbf{v}_s$ of the flowing screening current.}
\label{fig1}
\vspace{-0.4cm}
\end{figure}

Figure \ref{fig1} illustrates the claimed observation of the ``segmented Fermi surface'' of a $2$D Fermi gas under the influence of a superconducting order, when the surface state supports a steady screening current of velocity $\mathbf{v}_s$ driven by, for example, an external magnetic field.
A serious physical inconsistency can be observed via the corresponding electronic band energy-momentum $(\epsilon_\mathbf{k},\hbar\mathbf{k})$-dispersion in the right panel, when considering the average velocity $\braket{\hat{\mathbf{v}}} = (\hbar N)^{-1} \int f_\mathbf{k} \mathbf{\nabla}_\mathbf{k} \epsilon_\mathbf{k} d^2 k$ of the many-body surface state, with corresponding total particle number $N\equiv \int f_\mathbf{k} d^2 k$, occupying one-body momentum $\hbar\mathbf{k}$ with probability distribution $f_\mathbf{k}$. (Here $\hbar$ denotes the Planck constant.)
A surface state with the claimed Fermi surface corresponds to $f_\mathbf{k}\sim\theta(E_\mathrm{F}-\epsilon_\mathbf{k} )$~\cite{BogoliubovWeight}, i.e. fully occupied one-body states (in red) separated from the fully unoccupied ones by an overall Fermi level $E_\mathrm{F}$, assumed to coincide with the zero-bias voltage of the scanning tunneling spectroscopy (STS).
It is easy to visualize that the nearly even population of momentum $\hbar\mathbf{k}$ and $-\hbar\mathbf{k}$, highlighted by the thick red line on the axis, would dictate a nearly zero average momentum, \textit{inconsistent} with the finite flow.
Indeed, integration of the slope of $\epsilon_\mathbf{k}$ among the red occupied one-body states does not host the necessary finite flow velocity of the screening current ($\braket{\hat{\mathbf{v}}}\neq\mathbf{v}_s$)~\cite{AvgVel}.

To obtain a physically consistent picture, we first recall that all physical statistics, quantum or thermal, need to describe the same physical reality across different frames of observation. This fundamental requirement of the principle of relativity dictates that the probability distribution of states, such as $f_\mathbf{k}$, be inertial frame independent, even though observed properties can be quantitatively frame dependent. Therefore, the familiar form $e^{-\beta E}\rightarrow e^{-\beta \hat{H}}$ (with total energy $E$, Hamiltonian $\hat{H}$, and inverse energy scale of the temperature $\beta$) in general \emph{cannot} possibly give the physical Boltzmann relative probability, since neither $E$ or $\hat{H}$ is frame independent.
Instead, the Boltzmann relative probability $e^{-\beta E_0}$ \emph{requires} the frame-independent ``internal energy'' $E_0$ defined as the $E$ in the \emph{rest frame} of the system~\cite{landau5}. (Recall a similarly defined frame-independent scalar ``proper time'' $\tau$ in special theory of relativity.)

In other words, in an arbitrary inertial frame, the \textit{frame-independent} density matrix operator \textit{has} to be $e^{-\beta \hat{H}_0}=e^{-\beta \hat{U}^\dagger\hat{H}\hat{U}}$, where the \textit{frame-independent} internal energy operator $\hat{H}_0$ can be obtained from transforming the frame-dependent Hamiltonian $\hat{H}$ back to the rest frame with the help of the unitary boost operator $U$. (\textit{c.f.} Appendix II for a straightforward demonstration.)
Besides the desired frame independence, this identification also makes it easy to enforce the \emph{additional} (quantum/thermal) constraint that the observed average velocity of the system is correctly reproduced in any frame.
Correspondingly, all statistics related quantities in standard literature, such as Fermi surface, Fermi level $E_\mathrm{F}$, and chemical potential, can only be employed safely in the rest frame.
(\textit{c.f.} Appendix III for an example of the consequence due to frame mismatch between the frame-independent statistics controlled by $\hat{U}^\dagger\hat{H}\hat{U}$ and frame-dependent observables $\hat{O}$.)

\begin{figure}[t]
\renewcommand\thefigure{\arabic{figure}}
\includegraphics[height = 3.25cm]{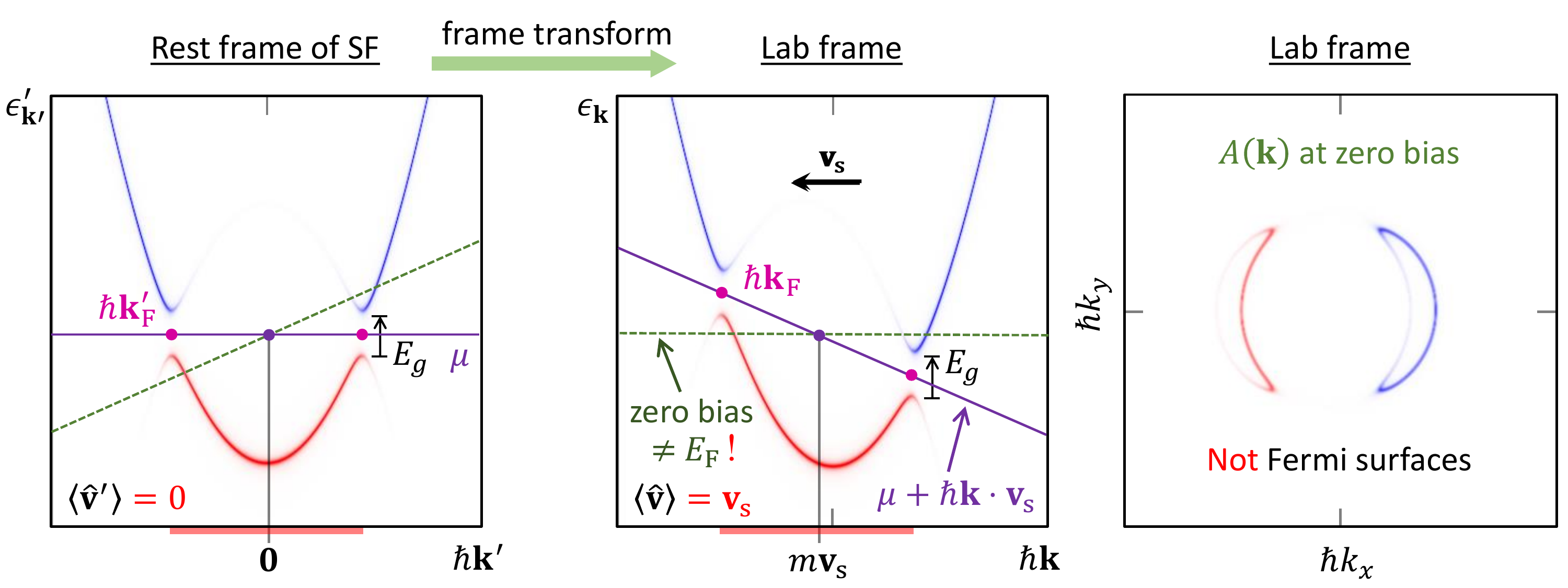}
\caption{Figure 2  Illustration of the energy-momentum $(\epsilon,\hbar\mathbf{k})$-dispersion of the one-body spectral function $A(\mathbf{k})$ for an $s$-wave superflow (SF) in its rest frame (left panel). Predominately occupied one-body states (in red) and predominately unoccupied ones (in blue) are fully separated by a superconductivity induced gap $E_g$ around the chemical potential $\mu$ (denoted by the purple dot). $\mathbf{k}_\mathrm{F}$ denotes the Fermi wavevector as part of the Fermi surface that only exists prior to the gap opening. Notice that in the lab frame (middle panel), each ($\epsilon_{\mathbf{k}}$, $\hbar\mathbf{k}$) is shifted by ($\hbar\mathbf{k}\cdot\mathbf{v}_s$, $m\mathbf{v}_s$), such that the momentum occupation is centered around $m\mathbf{v}_s$, as denoted by the thick red line along the $\hbar\mathbf{k}$-axis. Evidently, the zero-bias voltage of STS (in green) no longer separates the occupied and unoccupied one-body states. Instead, its interception with the dispersion only accesses Bogoliubov quasi-particle excitations originally with finite energy in the rest frame (c.f. left panel). The observed features at the zero-bias voltage (right panel) are therefore not Fermi surfaces, nor can they be zero modes such as the proposed Majorana end states.}
\label{fig2}
\vspace{-0.5cm}
\end{figure}

Following this basic requirement, Fig.\ref{fig2} illustrates the proper description of the above screening current in its rest frame and how it transforms to the lab frame. In the rest frame, a state experiencing an $s$-wave superconducting order has a full gap $E_g$ in the dispersion $\epsilon_{\mathbf{k}'}'$, in which the chemical potential $\mu$ corresponds to a $\mathbf{k}'$-independent energy level that separates the predominately occupied one-body state (in red) from those predominately unoccupied ones (in blue) through the Fermi-Dirac distribution $f_{\mathbf{k}'}=n_{\mathrm{F}} (\epsilon_{\mathbf{k}'}'-\mu)$ at finite temperature. Clearly, the average velocity $\braket{\hat{\mathbf{v}}}$ is indeed consistently zero.

The middle panel gives the \textit{same surface state} transformed to the lab frame, via for example a Galilean boost of space-time $(\mathbf{x}',t' )\rightarrow(\mathbf{x},t)=(\mathbf{x}'+ \mathbf{v}_s t',t' )$ with the velocity $\mathbf{v}_s$ of the screening current, and correspondingly $\epsilon_\mathbf{k}=\epsilon_{\mathbf{k}'}'+\hbar \mathbf{k}'\cdot\mathbf{v}_s$ and most essentially $f_\mathbf{k}=f_{\mathbf{k}'}$.  Using $d\mathbf{p}'=d\mathbf{p}=\hbar d\mathbf{k}$, it is straightforward to verify that the average velocity of the system is exactly $\mathbf{v}_s$, $\braket{\hat{\mathbf{v}}}=\braket{\hat{\mathbf{v}}'+\mathbf{v}_s}=\braket{\hat{\mathbf{v}}'}+\mathbf{v}_s=\mathbf{v}_s$.
Consistently, the momentum occupation is centered around $m\mathbf{v}_s$ as illustrated in the middle panel by the thick red line along the axis.
One thus obtains a physically consistent description through the use of the rest frame, while adhering to the fundamental principle of relativity.

This physically consistent description results in a completely different explanation of the experimental observation. As shown in the middle panel of Fig.\ref{fig2}, the zero-bias voltage (in green) of STS for such a flowing system no longer coincides with the $\mathbf{k}$-dependent purple line $\mu+\hbar\mathbf{k}\cdot\mathbf{v}_s$~\cite{mahan} that separates the occupied and unoccupied one-body states. Instead, the middle and right panels show that in general for a flowing system the zero bias of STS only accesses some of the occupied (in red) and unoccupied (in blue) quasiparticle excitations with \emph{finite} energy in the rest frame (c.f. left panel), as recently reaffirmed~\cite{davis1}. The experimentally observed features are therefore not Fermi surfaces, nor can they be zero modes such as the proposed Majorana end states.

Experimental verification of our counter-explanation should not be too difficult, given the dramatic difference between physical properties of gapless metals and those of fully gapped superconductors/insulators.
For example, compared to a metal with Fermi surface, a fully gapped superconductor would show highly suppressed entropy and thermal transport at low temperature.
Similarly, the optical conductivity should display a clean gap below the superconducting gap scale, as opposed to the Drude structure commonly observed in metals with Fermi surfaces.
Furthermore, the angular resolved photoemission spectroscopy would only observe the occupied part similar to the red band in the mid panel of Fig.~\ref{fig2} with a clean superconducting gap, but not any blue feature as implied by the existence of a segmented Fermi surface.
On the other hand, any measurements sensitive to phase-transition such as specific heat should show no sign of qualitative changes accompanying the occurrence of the claimed ``segmented Fermi surface'' upon increasing gradually the external magnetic field within the superconducting phase.

The most important aspect of the above rest-frame based description is its strict adherence to the principle of relativity. Similar to the above example, a gapped insulator in the rest frame would remain a gapped insulator in the lab frame as well, without metallic transport or any other long-wavelength low-energy excitation. Such a physically consistent description will not allow one observer's fully gapped $s$-wave superconductor to become gapless to another observer~\cite{jfjia}, or one observer's Mott insulator to become another observer's superconductor~\cite{ye1}. In fact, from the above consideration of physical statistics, it is straightforward to verify that these unphysical proposals originate from incorrect substitution of internal energy by total energy in an arbitrary frame~\cite{tinkham,volovik1,ye1}, or equivalently incorrect assignment of one-body occupation $f_\mathbf{k}$ (for example, via a single $\mathbf{k}$-independent energy level such as chemical potential or Fermi level~\cite{volovik2,fu1,fu2} outside the rest frame).

Note that the fundamental principle of relativity should apply to all frame transforms, Galilean or Lorentz, even though the convenient trick of employing the rest frame might not be always applicable, for example, when a single physically meaningful rest frame cannot be identified. Examples can include systems containing well-defined components of different mesoscopic flow velocities, ensembles of relativistic objects, and systems with strongly curved space time in general relativity. Regardless, as recurrently urged in the literature~\cite{pitaevskii1,AoThouZhu}, adherence to the principle of relativity is essential, even if this would require an extension of the current lore of physical statistics through further experimental investigation.

In summary, we advocate the necessity of the principle of relativity and demonstrate its consistent application through the use of the rest frame in defining proper physical statistics for flowing systems. Violation of this basic requirement resulted in several recent exotic claims that have received widespread attention. Using the recent claimed discovery of ``segmented Fermi surface'' as an example, we demonstrate that such a violation would necessarily lead to serious physical inconsistency between the observed and calculated average velocity. Upon a proper employment of the principle, we propose an alternative and physically consistent counterclaim and suggest means of experimental verification. This study showcases the importance of adherence to the basic principle of relative motion in physical statistics especially when pushing the frontiers of physics and technology.

\begin{acknowledgments}
We thank J.C. S\'eamus Davis, A.J. Leggett, G.A. Sawatzky, B.I. Halperin,  J. Schmalian, Ping Ao, and Weikang Lin for useful discussions. We give special thanks to Hao Zheng, Yongde Zhang, and Jianda Wu for the encouragement to publish this article. This work is supported by National Natural Science Foundation of China No. 12274287 and No. 12042507 and Innovation Program for Quantum Science and Technology No. 2021ZD0301900.
\end{acknowledgments}

\begin{center}

\textbf{\large Appendix I: Galilean boost of quasi-particles in one-body propagators}
\end{center}

In this appendix we illustrate that the frame transformation of quasiparticles, represented via one-body propagator, follows the standard Galilean boost. This closely parallels the well-known derivation commonly found in the literature~\cite{landau9}.

Consider, for example, an experimental process of inverse angle-resolved photoemission spectroscopy, in which one electron is injected into the sample and a photon is emitted.
Such process corresponds to the blue part of the quasiparticle dispersions illustrated in Fig.~\ref{fig1} and Fig.~\ref{fig2}.
It can be described by the imaginary part of the following contribution to the time-ordered one-body propagator with poles below the real frequency axis~\cite{negele-orland}
\begin{align}
G(\mathbf{k},\omega) &= \frac{\bra{\Psi_{N}}c_{\mathbf{k}} \ket{\Psi_{N+1}}\bra{\Psi_{N+1}} c^\dagger_{\mathbf{k}} \ket{\Psi_N}}{\omega - (E_{N+1}-E_{N})+i0^+}
\end{align}
of the system in a $N$-particle many-body state $\ket{\Psi_N}$ with corresponding total energy $E_{N}$ and total momentum $\hbar\mathbf{K}_N$.
The dispersion in those figures represents the energy-momentum $(\omega,\hbar\mathbf{k})=(E_{N+1}-E_{N},\hbar (\mathbf{K}_{N+1}-\mathbf{K}_N))$ required for such particle-addition excitations.
Therefore, the transformation of $(\omega,\hbar\mathbf{k})$ can be derived from the Galilean transformation of the \textit{total} energy-momentum $(E',\hbar\mathbf{K}')\rightarrow(E,\hbar\mathbf{K}))=(E'+\hbar\mathbf{K}\cdot\mathbf{u},\hbar\mathbf{K}'+M\mathbf{u})$, where $M$ is the total mass of the system,
\begin{align}
\label{GalE}
\omega&=E_{N+1}-E_N \nonumber \\
&= (E'_{N+1} - E'_N) + \hbar(\mathbf{K}'_{N+1}-\mathbf{K}'_N)\cdot\mathbf{u} \nonumber\\
&= \omega^\prime + \hbar\mathbf{k}'\cdot\mathbf{u}\\
\label{GalK}
\hbar\mathbf{k}&=\hbar(\mathbf{K}'_{N+1}-\mathbf{K}'_N)+(M_{N+1}-M_N)\mathbf{u} \nonumber\\
&=\hbar\mathbf{k}'+m\mathbf{u},
\end{align}
just like the standard transformation for classical particles.
Note that this fact is independent of the microscopic details generating the dispersion itself. In the present work, for example, the same transformation must hold for a pairing-generated superconductor.

\begin{center}

\textbf{\large Appendix II: Galilean boost of thermal statistics for observables}
\end{center}

This appendix provides a pedagogical demonstration for the \textit{proper} (frame-independent) density matrix in an arbitrary inertial frame.
Consider, for example, a system in its rest frame in the grand canonical ensemble with corresponding density matrix $\hat{\rho} \equiv e^{-\beta(\hat{H}'-\mu \hat{N}')}$  determined by the Hamiltonian $\hat{H}'$ in \textit{this} frame, whose expectation value gives the frame-independent \textit{proper} internal energy.
The thermal average of an observable $\hat{O}'$ in that rest frame can be represented by eigenstates $\ket{I'}$ of $\hat{H}'$,
\begin{align}
\nonumber
\braket{\hat{O}'} &= \frac{1}{Z}\mathrm{Tr}\Big( e^{-\beta (\hat{H}'-\mu \hat{N}')} \hat{O}' \Big) \\ \nonumber
&= \frac{1}{Z}\sum_{I'} \bra{I'} e^{-\beta (\hat{H}'-\mu \hat{N}')} \hat{O}' \ket{I'}\\
&= \frac{1}{Z}\sum_{I'} e^{-\beta (E_{I'}-\mu N_{I'})} \bra{I'} \hat{O}' \ket{I'}\\
&= \sum_{I'} P(E_{I'}, N_{I'}) \bra{I'} \hat{O}' \ket{I'}
\label{FETransf1}
\end{align}
with corresponding eigenvalues $E_I'$ and particle numbers $N_I'$.
Here the partition function $Z=\mathrm{Tr}\Big( e^{-\beta (\hat{H}'-\mu \hat{N}')} \Big)$ properly normalizes the probability $P(E_{I'}, N_{I'})$.
Since the probability is frame independent, a Galilean boost would only alter the quantum expectation value of the observable via a unitary boost operator $\hat{U}^{\dag}$,
\begin{align}
\bra{I'} \hat{O}' \ket{I'} \rightarrow \bra{I'} \hat{U} \hat{O}' \hat{U}^{\dag} \ket{I'} = \bra{I'} \hat{O} \ket{I'},
\end{align}
where $\hat{O}$ is the observable operator in the boost frame.
The resulting thermal average in this frame then becomes
\begin{align}
\nonumber
\braket{\hat{O}} &= \sum_{I'} P(E_{I'}, N_{I'}) \bra{I'} \hat{U} \hat{O}' \hat{U}^{\dag} \ket{I'}\\
&= \frac{1}{Z}\sum_{I'} e^{-\beta (E_{I'}-\mu N_{I'})} \bra{I'} \hat{O} \ket{I'} \nonumber \\
&= \frac{1}{Z}\mathrm{Tr}\Big( e^{-\beta (\hat{H}'-\mu \hat{N}')} \hat{O} \Big) \nonumber \\
&= \frac{1}{Z}\mathrm{Tr}\Big( e^{-\beta (\hat{U}^{\dag}\hat{H}\hat{U}-\mu \hat{N})} \hat{O} \Big),
\label{FETransf2}
\end{align}
where the frame invariance of total particle number $\hat{N}$ is employed, and the frame-independent internal energy operator $\hat{H}'$ is obtained via inverse transform of the frame-dependent Hamiltonian $\hat{H}$ back to the rest frame.
The proper density matrix $\hat{\rho}$ in an arbitrary frame is therefore
\begin{align}
\hat{\rho}&=e^{-\beta (\hat{U}^{\dag}\hat{H}\hat{U}-\mu \hat{N})}~~\Big(=e^{-\beta (\hat{H}'-\mu \hat{N}')}\Big)\\
\nonumber
&\neq e^{-\beta (\hat{H}-\mu \hat{N})},
\end{align}
as obviously
$\frac{1}{Z}\mathrm{Tr}\Big( e^{-\beta (\hat{U}^{\dag}\hat{H}\hat{U}-\mu \hat{N})} \hat{O} \Big) \neq \frac{1}{Z}\mathrm{Tr}\Big( e^{-\beta (\hat{H}-\mu \hat{N})} \hat{O} \Big)$.
\begin{center}

\textbf{\large Appendix III: An example of applying Galilean boost to a superconducting state}
\end{center}

In this appendix we construct an explicit example of frame transformation in the context of superconductivity. The essence of such a transformation is that physically there should be only one reality. Correspondingly, a change of frame can only \textit{quantitatively} change the observables of the states, but not their qualitative nature. 

To demonstrate the necessity of tracking the change of observables of a state across frames, here we illustrate the consequences of the (Galilean) principle of relative motion on the ground state of the well-known Bogoliubov-de Genne (BdG) mean-field treatment~\cite{altland-simons}. In the rest frame of a superconductor, the ground state is identified via
\begin{align}
H-\mu N &= \sum_{\mathbf{p'}\sigma} \Omega_\mathbf{p'} d^{\dag}_{\mathbf{p'}\sigma} d_{\mathbf{p'}\sigma} \\
\begin{pmatrix}
d_{\mathbf{p'}\uparrow} \\
d^{\dag}_{-\mathbf{p'}\downarrow}
\end{pmatrix} &=
\begin{pmatrix}
\cos{\theta_{\mathbf{p'}}} & \sin{\theta_{\mathbf{p'}}} \\
\sin{\theta_{\mathbf{p'}}} & -\cos{\theta_{\mathbf{p'}}}
\end{pmatrix}
\begin{pmatrix}
c_{\mathbf{p'}\uparrow} \\
c^{\dag}_{-\mathbf{p'}\downarrow}
\end{pmatrix} \\
\sin \theta_{\mathbf{p'}} &= \sqrt{\frac{1}{2}\Big( 1 - \frac{\epsilon_{\mathbf{p'}}-\mu}{\Omega_{\mathbf{p'}}} \Big)} \\
\Omega_{\mathbf{p'}} &= \sqrt{\Delta^2+(\epsilon_{\mathbf{p'}}-\mu)^2},
\end{align}
with quasiparticle annihilation operator $d_{\mathbf{p'}\sigma}$ for spin $\sigma\in\{\uparrow,\downarrow\}$, quasiparticle dispersion $\Omega_{\mathbf{p'}}$, self-consistently determined superconducting gap $\Delta$, and bare dispersion $\epsilon_{\mathbf{p'}}$. The ground state $\ket{G}'$ is then given by
\begin{align}
\label{RestMBGS}
\ket{G}' &= \prod_{\mathbf{p}'} d_{\mathbf{p'}\uparrow} d_{-\mathbf{p'}\downarrow} \ket{0} = \prod_{\mathbf{p}'} \Big( \cos{\theta_{\mathbf{p'}}} - \sin{\theta_{\mathbf{p'}}}c^{\dag}_{\mathbf{p'}\uparrow} c^{\dag}_{-\mathbf{p'}\downarrow} \Big) \ket{0},
\end{align}
where $\ket{0}$ is the true vacuum with no electrons.

Now, to describe the same state in another frame, we carefully choose a unitary operator that results in a Galilean transformation. One such operator is
\begin{align}
\label{GalU}
U &= e^{i\hat{\mathbf{p}} \cdot \mathbf{u}t / \hbar} e^{-im\mathbf{u} \cdot \hat{\mathbf{x}} / \hbar},
\end{align}
which can easily be verified to produce the following transformations on the \textit{basis} of second quantization, namely the creation and annihilation operators,
\begin{align}
\label{cx}
\tilde{c}^{\dag}_{\mathbf{x}\uparrow} \equiv U c^{\dag}_{\mathbf{x}\uparrow} U^{\dag} &= e^{-im\mathbf{u} \cdot \mathbf{x} / \hbar} c^{\dag}_{\mathbf{x}-\mathbf{u}t\uparrow} \\
\label{ck}
\tilde{c}^{\dag}_{\mathbf{p}\uparrow} \equiv U c^{\dag}_{\mathbf{p}\uparrow} U^{\dag} &=e^{-i(\mathbf{p}-m\mathbf{u}/ \hbar) \cdot \mathbf{u}t} c^{\dag}_{\mathbf{p}-m\mathbf{u}\uparrow}.
\end{align}

Correspondingly, the second-quantized \textit{operators}, such as position $\hat{\mathbf{x}}$ and momentum $\hat{\mathbf{p}}$, transform~\cite{StatevOp} via
\begin{align}
\label{xTransf}
\hat{\mathbf{x}}^\prime &= U^{\dag} \hat{\mathbf{x}} U
= U^{\dag} \int d \mathbf{x} \sum_{\sigma} \mathbf{x} c^{\dag}_{\mathbf{x}\sigma} c_{\mathbf{x}\sigma} U\\ \nonumber
&= \int d \mathbf{x} \sum_{\sigma} \mathbf{x} c^{\dag}_{\mathbf{x}+\mathbf{u}t \sigma} c_{\mathbf{x}+\mathbf{u}t \sigma} = \int d \mathbf{x} \sum_{\sigma} (\mathbf{x}-\mathbf{u} t) c^{\dag}_{\mathbf{x}\sigma} c_{\mathbf{x}\sigma}\\ \nonumber
&= \hat{\mathbf{x}}-\mathbf{u}t \\
\label{kTransf}
\hat{\mathbf{p}}^\prime &= U^{\dag} \hat{\mathbf{p}} U = U^{\dag} \int d \mathbf{p} \sum_{\sigma} \mathbf{p} c^{\dag}_{\mathbf{p}\sigma} c_{\mathbf{p}\sigma} U\\ \nonumber
&= \int d \mathbf{p} \sum_{\sigma} \mathbf{p} c^{\dag}_{\mathbf{p}+m\mathbf{u} \sigma} c_{\mathbf{p}+m\mathbf{u} \sigma}
= \int d \mathbf{p} \sum_{\sigma} (\mathbf{p}-m\mathbf{u}) c^{\dag}_{\mathbf{p}\sigma} c_{\mathbf{p}\sigma}\\ \nonumber
&= \hat{\mathbf{p}}-m\mathbf{u},\nonumber
\end{align}
as required by Galilean transformation.

The ground state $\ket{G}'$ is then described in the lab frame via Eqs.(\ref{cx}) and (\ref{ck}),
\begin{align}
\label{GalMBGS}
\ket{G} &= U^{\dag}\ket{G}'\\
&= \prod_{\mathbf{p}'} \Big( \cos{\theta_{\mathbf{p'}}} - \sin{\theta_{\mathbf{p'}}} e^{2imu^2t/\hbar} c^{\dag}_{\mathbf{p'}+m\mathbf{u}\uparrow} c^{\dag}_{-\mathbf{p'}+m\mathbf{u}\downarrow} \Big) \ket{0},\nonumber
\end{align}
where $\mathbf{u}=\mathbf{v}_s$ is now the velocity of the superflow.
Not surprisingly, states are \textit{not} invariant under frame transformations. Note, however, a non-trivial additional $2mu^2t/\hbar$ relative phase appears in $\ket{G}$, reflecting the spontaneous global $U(1)$ symmetry breaking in superconducting systems.

More relevant to the main point of this manuscript, concerning the physical statistics (or probability), in great contrast to $\ket{G}'$, $\ket{G}$ in the lab frame has different occupation between $\mathbf{p}$ and $-\mathbf{p}$.
Indeed, according to Eq.~(\ref{GalMBGS}) occupation of $\mathbf{p}=m\mathbf{v}_s+\mathbf{p'}$ is the same as that of $m\mathbf{v}_s-\mathbf{p'}$ (but \textit{not} $-\mathbf{p}=-m\mathbf{v}_s-\mathbf{p'}$).
This immediately confirms the physical inconsistency as presented in Fig.~\ref{fig1}, namely a steady flow with a finite velocity $\mathbf{v}_s$ cannot be consistently described by an energy-only dependent occupation assignment, e.g. $f_\mathbf{k}\sim\theta(E_\mathrm{F}-\epsilon_\mathbf{k} )$, as the latter would result in a nearly symmetric occupation between $\mathbf{p}$ and $-\mathbf{p}$ and correspondingly an average velocity unrelated to $\mathbf{v}_s$.

Furthermore, Eq.~(\ref{GalMBGS}) also shows that the electronic occupation of the system is centered around momentum $m\mathbf{v}_s$, confirming the picture presented in Fig.~\ref{fig2}.
Correspondingly, the occupation cannot be described by a simple energy-only occupation function in the lab frame.
Therefore, as already indicated in Appendix II by the frame mismatch between the frame-independent statistics controlled by $\hat{U}^\dagger\hat{H}\hat{U}$ and frame-dependent observables $\hat{O}$, for systems with steady flow velocity, assignment of a single energy level as a Fermi energy or chemical potential is not possible.

\typeout{get arXiv to do 4 passes: Label(s) may have changed. Rerun}


\begin{thebibliography}{20}%
\makeatletter
\providecommand \@ifxundefined [1]{%
 \@ifx{#1\undefined}
}%
\providecommand \@ifnum [1]{%
 \ifnum #1\expandafter \@firstoftwo
 \else \expandafter \@secondoftwo
 \fi
}%
\providecommand \@ifx [1]{%
 \ifx #1\expandafter \@firstoftwo
 \else \expandafter \@secondoftwo
 \fi
}%
\providecommand \natexlab [1]{#1}%
\providecommand \enquote  [1]{``#1''}%
\providecommand \bibnamefont  [1]{#1}%
\providecommand \bibfnamefont [1]{#1}%
\providecommand \citenamefont [1]{#1}%
\providecommand \href@noop [0]{\@secondoftwo}%
\providecommand \href [0]{\begingroup \@sanitize@url \@href}%
\providecommand \@href[1]{\@@startlink{#1}\@@href}%
\providecommand \@@href[1]{\endgroup#1\@@endlink}%
\providecommand \@sanitize@url [0]{\catcode `\\12\catcode `\$12\catcode
  `\&12\catcode `\#12\catcode `\^12\catcode `\_12\catcode `\%12\relax}%
\providecommand \@@startlink[1]{}%
\providecommand \@@endlink[0]{}%
\providecommand \url  [0]{\begingroup\@sanitize@url \@url }%
\providecommand \@url [1]{\endgroup\@href {#1}{\urlprefix }}%
\providecommand \urlprefix  [0]{URL }%
\providecommand \Eprint [0]{\href }%
\providecommand \doibase [0]{http://dx.doi.org/}%
\providecommand \selectlanguage [0]{\@gobble}%
\providecommand \bibinfo  [0]{\@secondoftwo}%
\providecommand \bibfield  [0]{\@secondoftwo}%
\providecommand \translation [1]{[#1]}%
\providecommand \BibitemOpen [0]{}%
\providecommand \bibitemStop [0]{}%
\providecommand \bibitemNoStop [0]{.\EOS\space}%
\providecommand \EOS [0]{\spacefactor3000\relax}%
\providecommand \BibitemShut  [1]{\csname bibitem#1\endcsname}%
\let\auto@bib@innerbib\@empty
\bibitem [{\citenamefont {Zhu}\ \emph {et~al.}(2021)\citenamefont {Zhu},
  \citenamefont {Papaj}, \citenamefont {Nie}, \citenamefont {Xu}, \citenamefont
  {Gu}, \citenamefont {Yang}, \citenamefont {Guan}, \citenamefont {Wang},
  \citenamefont {Li}, \citenamefont {Liu}, \citenamefont {Luo}, \citenamefont
  {Xu}, \citenamefont {Zheng}, \citenamefont {Fu},\ and\ \citenamefont
  {Jia}}]{jfjia}%
  \BibitemOpen
  \bibfield  {author} {\bibinfo {author} {\bibfnamefont {Z.}~\bibnamefont
  {Zhu}}, \bibinfo {author} {\bibfnamefont {M.}~\bibnamefont {Papaj}}, \bibinfo
  {author} {\bibfnamefont {X.-A.}\ \bibnamefont {Nie}}, \bibinfo {author}
  {\bibfnamefont {H.-K.}\ \bibnamefont {Xu}}, \bibinfo {author} {\bibfnamefont
  {Y.-S.}\ \bibnamefont {Gu}}, \bibinfo {author} {\bibfnamefont
  {X.}~\bibnamefont {Yang}}, \bibinfo {author} {\bibfnamefont {D.}~\bibnamefont
  {Guan}}, \bibinfo {author} {\bibfnamefont {S.}~\bibnamefont {Wang}}, \bibinfo
  {author} {\bibfnamefont {Y.}~\bibnamefont {Li}}, \bibinfo {author}
  {\bibfnamefont {C.}~\bibnamefont {Liu}}, \bibinfo {author} {\bibfnamefont
  {J.}~\bibnamefont {Luo}}, \bibinfo {author} {\bibfnamefont {Z.-A.}\
  \bibnamefont {Xu}}, \bibinfo {author} {\bibfnamefont {H.}~\bibnamefont
  {Zheng}}, \bibinfo {author} {\bibfnamefont {L.}~\bibnamefont {Fu}}, \ and\
  \bibinfo {author} {\bibfnamefont {J.-F.}\ \bibnamefont {Jia}},\ }\href@noop
  {} {\bibfield  {journal} {\bibinfo  {journal} {Science}\ }\textbf {\bibinfo
  {volume} {374}},\ \bibinfo {pages} {1381} (\bibinfo {year}
  {2021})}\BibitemShut {NoStop}%
\bibitem [{\citenamefont {Sun}\ and\ \citenamefont {Ye}(2023)}]{ye1}%
  \BibitemOpen
  \bibfield  {author} {\bibinfo {author} {\bibfnamefont {F.}~\bibnamefont
  {Sun}}\ and\ \bibinfo {author} {\bibfnamefont {J.}~\bibnamefont {Ye}},\
  }\href@noop {} {\  (\bibinfo {year} {2023})},\ \Eprint
  {http://arxiv.org/abs/2207.10475} {arXiv:2207.10475 [cond-mat.str-el]}
  \BibitemShut {NoStop}%
\bibitem [{\citenamefont {London}\ and\ \citenamefont {London}(1935)}]{london}%
  \BibitemOpen
  \bibfield  {author} {\bibinfo {author} {\bibfnamefont {H.}~\bibnamefont
  {London}}\ and\ \bibinfo {author} {\bibfnamefont {F.}~\bibnamefont
  {London}},\ }\href@noop {} {\bibfield  {journal} {\bibinfo  {journal}
  {Proceedings of the Royal Society A}\ }\textbf {\bibinfo {volume} {149}},\
  \bibinfo {pages} {71} (\bibinfo {year} {1935})}\BibitemShut {NoStop}%
\bibitem [{\citenamefont {Landau}\ and\ \citenamefont
  {Lifshitz}(1980{\natexlab{a}})}]{landau9}%
  \BibitemOpen
  \bibfield  {author} {\bibinfo {author} {\bibfnamefont {L.}~\bibnamefont
  {Landau}}\ and\ \bibinfo {author} {\bibfnamefont {E.}~\bibnamefont
  {Lifshitz}},\ }\href@noop {} {\emph {\bibinfo {title} {Statistical
  Physics}}},\ \bibinfo {edition} {2nd}\ ed.,\ Vol.~\bibinfo {volume} {9}\
  (\bibinfo  {publisher} {Pergamon Press},\ \bibinfo {year} {1980})\ pp.\
  \bibinfo {pages} {88--91}\BibitemShut {NoStop}%
\bibitem [{Bog()}]{BogoliubovWeight}%
  \BibitemOpen
  \href@noop {} {}\bibinfo {note} {The more precise expression,
  $f_\mathbf{k}=w(\epsilon_\mathbf{k})\times\theta({E}_\textrm{F}-\epsilon_\mathbf{k})$,
  includes minor $\mathbf{v}_s$-\textit{independent} correction,
  $w(\epsilon_\mathbf{k})=\frac{1}{2}(1-\tfrac{\epsilon_\mathbf{k}}{\sqrt{\epsilon_\mathbf{k}^2+{\Delta}^2}})$,
  to account for the mixed particle-hole nature of Bogoliubov quasiparticles
  near the superconducting gap ${\Delta}$.}\BibitemShut {Stop}%
\bibitem [{Avg()}]{AvgVel}%
  \BibitemOpen
  \href@noop {} {}\bibinfo {note} {Instead, it has a nearly zero average
  velocity $\braket{\hat{\mathbf{v}}}$ that reflects the small
  $\mathbf{v}_s$-\textit{independent} gap opening.}\BibitemShut {Stop}%
\bibitem [{\citenamefont {Landau}\ and\ \citenamefont
  {Lifshitz}(1980{\natexlab{b}})}]{landau5}%
  \BibitemOpen
  \bibfield  {author} {\bibinfo {author} {\bibfnamefont {L.}~\bibnamefont
  {Landau}}\ and\ \bibinfo {author} {\bibfnamefont {E.}~\bibnamefont
  {Lifshitz}},\ }\href@noop {} {\emph {\bibinfo {title} {Statistical
  Physics}}},\ \bibinfo {edition} {3rd}\ ed.,\ Vol.~\bibinfo {volume} {5}\
  (\bibinfo  {publisher} {Pergamon Press},\ \bibinfo {year} {1980})\
  p.~\bibinfo {pages} {36}\BibitemShut {NoStop}%
\bibitem [{\citenamefont {Mahan}(2000)}]{mahan}%
  \BibitemOpen
  \bibfield  {author} {\bibinfo {author} {\bibfnamefont {G.}~\bibnamefont
  {Mahan}},\ }\href@noop {} {\emph {\bibinfo {title} {Many-Particle
  Physics}}},\ \bibinfo {edition} {3rd}\ ed.\ (\bibinfo  {publisher} {Springer
  New York},\ \bibinfo {year} {2000})\ pp.\ \bibinfo {pages}
  {499--501}\BibitemShut {NoStop}%
\bibitem [{\citenamefont {{Liu, X. and Chong, Y.X. and Sharma, R. and Davis,
  J.C.S.}}(2021)}]{davis1}%
  \BibitemOpen
  \bibfield  {author} {\bibinfo {author} {\bibnamefont {{Liu, X. and Chong,
  Y.X. and Sharma, R. and Davis, J.C.S.}}},\ }\href@noop {} {\bibfield
  {journal} {\bibinfo  {journal} {Nature Materials}\ }\textbf {\bibinfo
  {volume} {20}},\ \bibinfo {pages} {1480} (\bibinfo {year}
  {2021})}\BibitemShut {NoStop}%
\bibitem [{\citenamefont {Tinkham}(1996)}]{tinkham}%
  \BibitemOpen
  \bibfield  {author} {\bibinfo {author} {\bibfnamefont {M.}~\bibnamefont
  {Tinkham}},\ }\href@noop {} {\emph {\bibinfo {title} {Introduction to
  Superconductivity}}}\ (\bibinfo  {publisher} {McGraw-Hill},\ \bibinfo {year}
  {1996})\ pp.\ \bibinfo {pages} {387--388}\BibitemShut {NoStop}%
\bibitem [{\citenamefont {Volovik}(2003)}]{volovik1}%
  \BibitemOpen
  \bibfield  {author} {\bibinfo {author} {\bibfnamefont {G.}~\bibnamefont
  {Volovik}},\ }\href@noop {} {\emph {\bibinfo {title} {The Universe in a
  Helium Droplet}}}\ (\bibinfo  {publisher} {Oxford University Press},\
  \bibinfo {year} {2003})\ pp.\ \bibinfo {pages} {323--324}\BibitemShut
  {NoStop}%
\bibitem [{\citenamefont {Volovik}(2007)}]{volovik2}%
  \BibitemOpen
  \bibfield  {author} {\bibinfo {author} {\bibfnamefont {G.}~\bibnamefont
  {Volovik}},\ }\href@noop {} {\emph {\bibinfo {title} {Unruh, W.G.,
  Sch\"{u}tzhold, R. (eds) Quantum Analogues: From Phase Transitions to Black
  Holes and Cosmology}}},\ Vol.\ \bibinfo {volume} {718}\ (\bibinfo
  {publisher} {Springer, Berlin, Heidelberg},\ \bibinfo {year} {2007})\ Chap.\
  \bibinfo {chapter} {Quantum Phase Transitions from Topology in Momentum
  Space}, p.~\bibinfo {pages} {7}\BibitemShut {NoStop}%
\bibitem [{\citenamefont {Yuan}\ and\ \citenamefont {Fu}(2018)}]{fu1}%
  \BibitemOpen
  \bibfield  {author} {\bibinfo {author} {\bibfnamefont {N.~F.~Q.}\
  \bibnamefont {Yuan}}\ and\ \bibinfo {author} {\bibfnamefont {L.}~\bibnamefont
  {Fu}},\ }\href@noop {} {\bibfield  {journal} {\bibinfo  {journal} {Phys. Rev.
  B}\ }\textbf {\bibinfo {volume} {97}},\ \bibinfo {pages} {115139} (\bibinfo
  {year} {2018})}\BibitemShut {NoStop}%
\bibitem [{\citenamefont {Papaj}\ and\ \citenamefont {Fu}(2021)}]{fu2}%
  \BibitemOpen
  \bibfield  {author} {\bibinfo {author} {\bibfnamefont {M.}~\bibnamefont
  {Papaj}}\ and\ \bibinfo {author} {\bibfnamefont {L.}~\bibnamefont {Fu}},\
  }\href@noop {} {\bibfield  {journal} {\bibinfo  {journal} {Nature
  Communications}\ }\textbf {\bibinfo {volume} {12}},\ \bibinfo {pages} {577}
  (\bibinfo {year} {2021})}\BibitemShut {NoStop}%
\bibitem [{\citenamefont {Kemoklidze}\ and\ \citenamefont
  {Pitaevskii}(1966)}]{pitaevskii1}%
  \BibitemOpen
  \bibfield  {author} {\bibinfo {author} {\bibfnamefont {M.}~\bibnamefont
  {Kemoklidze}}\ and\ \bibinfo {author} {\bibfnamefont {L.}~\bibnamefont
  {Pitaevskii}},\ }\href@noop {} {\bibfield  {journal} {\bibinfo  {journal}
  {Soviet Physics JETP}\ }\textbf {\bibinfo {volume} {23}},\ \bibinfo {pages}
  {160} (\bibinfo {year} {1966})}\BibitemShut {NoStop}%
\bibitem [{\citenamefont {Aitchison}\ \emph {et~al.}(1995)\citenamefont
  {Aitchison}, \citenamefont {Ao}, \citenamefont {Thouless},\ and\
  \citenamefont {Zhu}}]{AoThouZhu}%
  \BibitemOpen
  \bibfield  {author} {\bibinfo {author} {\bibfnamefont {I.~J.~R.}\
  \bibnamefont {Aitchison}}, \bibinfo {author} {\bibfnamefont {P.}~\bibnamefont
  {Ao}}, \bibinfo {author} {\bibfnamefont {D.~J.}\ \bibnamefont {Thouless}}, \
  and\ \bibinfo {author} {\bibfnamefont {X.-M.}\ \bibnamefont {Zhu}},\
  }\href@noop {} {\bibfield  {journal} {\bibinfo  {journal} {Phys. Rev. B}\
  }\textbf {\bibinfo {volume} {51}},\ \bibinfo {pages} {6531} (\bibinfo {year}
  {1995})}\BibitemShut {NoStop}%
\bibitem [{\citenamefont {Negele}\ and\ \citenamefont
  {Orland}(1998)}]{negele-orland}%
  \BibitemOpen
  \bibfield  {author} {\bibinfo {author} {\bibfnamefont {J.}~\bibnamefont
  {Negele}}\ and\ \bibinfo {author} {\bibfnamefont {H.}~\bibnamefont
  {Orland}},\ }\href@noop {} {\emph {\bibinfo {title} {Quantum Many-Particle
  Systems}}}\ (\bibinfo  {publisher} {Westview Press},\ \bibinfo {year}
  {1998})\ pp.\ \bibinfo {pages} {241--242}\BibitemShut {NoStop}%
\bibitem [{\citenamefont {Altland}\ and\ \citenamefont
  {Simons}(2006)}]{altland-simons}%
  \BibitemOpen
  \bibfield  {author} {\bibinfo {author} {\bibfnamefont {A.}~\bibnamefont
  {Altland}}\ and\ \bibinfo {author} {\bibfnamefont {B.}~\bibnamefont
  {Simons}},\ }\href@noop {} {\emph {\bibinfo {title} {Condensed Matter Field
  Theory}}}\ (\bibinfo  {publisher} {Cambridge University Press},\ \bibinfo
  {year} {2006})\ pp.\ \bibinfo {pages} {276--278}\BibitemShut {NoStop}%
\bibitem [{Sta()}]{StatevOp}%
  \BibitemOpen
  \href@noop {} {}\bibinfo {note} {Notice that the transformation of the
  operators is the inverse of that of the basis, analogous to the difference in
  the Heisenberg and Schr\"{o}dinger pictures~\cite{sakurai}.}\BibitemShut
  {Stop}%
\bibitem [{\citenamefont {Sakurai}(1994)}]{sakurai}%
  \BibitemOpen
  \bibfield  {author} {\bibinfo {author} {\bibfnamefont {J.}~\bibnamefont
  {Sakurai}},\ }\href@noop {} {\emph {\bibinfo {title} {Modern Quantum
  Mechanics}}}\ (\bibinfo  {publisher} {Addison-Wesely},\ \bibinfo {year}
  {1994})\ pp.\ \bibinfo {pages} {80--81}\BibitemShut {NoStop}%
\end{thebibliography}
%


\end{document}